\documentclass[runningheads]{llncs}
\usepackage[T1]{fontenc}
\usepackage{graphicx}
\usepackage{utfsym}
\usepackage{cite}
\usepackage[hidelinks,bookmarksopen=true]{hyperref}
\usepackage[T1]{fontenc}
\usepackage{graphicx}
\usepackage{cite}
\usepackage[hidelinks,bookmarksopen=true]{hyperref}
\begin{document}
\title{CHaystack: Benchmarking Chinese Document Retrieval and VQA}
%
%
\author{Hanxi Li\inst{1}}
\authorrunning{H. Li}
\institute{Sichuan University, Chengdu, China\\
\email{hanncie@outlook.com}}

\maketitle              
\begin{abstract}
Retrieval-augmented generation (RAG) has made substantial progress in extending the memory of large language models (LLMs), and recent advances have further pushed RAG from pure text settings toward multimodal scenarios. In the document understanding domain, document visual question answering (DocumentVQA) has evolved from question answering over a single document to retrieval-and-generation pipelines over large-scale document collections. However, a benchmark specifically designed for Chinese large-scale document retrieval and question answering is still lacking. To bridge this gap, we introduce CHaystack, a new Chinese DocumentVQA benchmark that covers four document categories, namely academic papers, advertisements, web pages, and real-world photographed documents, enabling a more comprehensive evaluation of DocumentVQA systems. In addition, we present CDocRAG, a Chinese DocumentVQA system that uses a VLM-based relevance filter to verify retrieved document images before answer generation. We evaluate representative open-source embedding and generation models on CHaystack. The results reveal a clear contrast in category-wise strengths: Qwen-family models perform best on text-rich documents such as webpages and papers, whereas other models only achieve competitive results on visually rich categories such as advertisements and degrade sharply on text-dense documents. For retrieval, Qwen3-VL reaches 71.91 Recall@1 while the best non-Qwen model achieves only 14.40. These results indicate that the core challenge of CHaystack lies in Chinese textual encoding, and that Chinese large-scale DocumentVQA still leaves substantial room for improvement. Our code and dateset is available at https://github.com/hanxi19/CHaystack

\end{abstract}

\section{Introduction}

Retrieval-augmented generation (RAG) has substantially extended the memory and knowledge access
capabilities of large language models (LLMs)~\cite{rag_original}. As RAG continues to evolve, recent       
studies have progressively extended it beyond pure-text settings into multimodal scenarios, where models
must jointly retrieve and reason over visual and textual evidence~\cite{zhang_comprehensive_2024}. Within  
this landscape, DocumentVQA has emerged as a particularly active research area, as it demands that models
jointly understand document structure, recognize text, and integrate visual and semantic cues for question
answering.

In this domain, DocumentVQA systems have evolved from question answering over a single                     
document~\cite{docvqa} to RAG pipelines operating over large-scale document
collections, as exemplified by recent systems such as Document                                 
Haystacks~\cite{chen_document_2024}. Despite this rapid progress, benchmarks and systems        
specifically designed for Chinese remain scarce. Existing Chinese benchmarks, such as DuReader and
MosaicDoc~\cite{qi_textrmdureader_textrmvis_2022, mosaicdoc}, remain limited in document diversity,
focusing primarily on web pages and newspapers, respectively. As a result, they do not fully reflect the
broad range of challenges encountered in real-world Chinese document understanding.

\begin{figure}[t]
        \centering
        \includegraphics[width=0.95\textwidth]{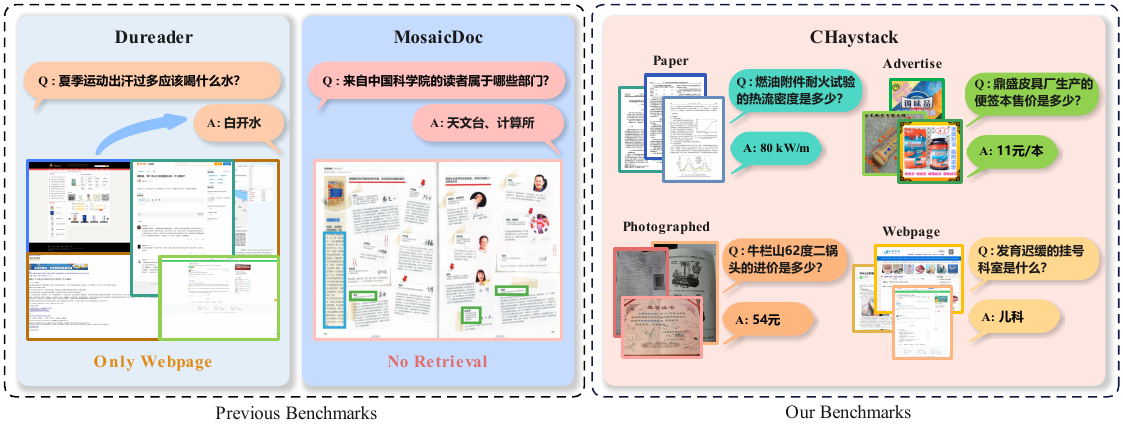}
        \caption{\textbf{Comparison between previous and our benchmarks.} Existing Chinese DocumentVQA benchmarks remain limited in either document diversity or the document retrieval. DuReader mainly focuses on webpage content, while MosaicDoc only contains newspapers and magazines and does not evaluate retrieval over a large candidate pool. In contrast, CHaystack organizes each question over a more diverse document collection spanning four categories, including academic papers, advertisements, photographed documents, and webpages. This setting better reflects real-world Chinese large-scale document retrieval and question answering, and provides a more challenging testbed for both retrieval accuracy and answer generation.}
        \label{fig:system}
\end{figure}

To bridge this gap, we introduce \textbf{CHaystack}, a new benchmark for Chinese large-scale DocumentVQA.  
As illustrated in Figure~\ref{fig:system}, CHaystack consists of four document categories, namely academic
papers, advertisements, web pages, and real-world photographed documents, thereby enabling a more          
comprehensive evaluation of system capabilities. Academic papers are typically text-rich and structurally
regular, placing strong demands on OCR quality and text understanding. Web pages usually contain a
relatively balanced mix of textual and visual elements together with abundant distracting content, which
makes localizing relevant evidence more difficult. Advertisements are more visually oriented and often
contain sparse text, requiring models to combine limited textual cues with layout and image semantics.
Real-world photographed documents are captured in unconstrained environments and therefore often suffer
from lighting variation, viewing angle changes, blur, and background clutter, all of which pose challenges
for robust evidence extraction.

We further propose \textbf{CDocRAG}, a Chinese DocumentVQA system for
large-scale retrieval and generation. Its key design is to insert a VLM-based relevance filter between
retrieval and generation. Instead of directly feeding the top-ranked retrieved images to the generator,
CDocRAG first asks a VLM to judge whether each candidate image contains evidence for the question, and then
uses the filtered candidates for answer generation. This verification step helps remove distracting
documents before generation and improves the quality of the evidence passed to the final VQA model.

We evaluate representative open-source retrieval and generation models on CHaystack. The results reveal distinct category preferences across models: Qwen-family models perform best on text-rich categories such as webpages and papers, whereas CLIP-family models favor visually rich categories such as advertisements. Despite this divergence, Qwen-family models attain substantially stronger overall performance, with Qwen3-VL reaching 71.91 Recall@1 for retrieval compared with 14.40 for the best non-Qwen model. These findings demonstrate that Chinese large-scale document retrieval and question answering still leave substantial room for improvement.
                                
Our contributions are summarized as follows:
\begin{itemize}                                                                                            
\item We introduce CHaystack, a new benchmark for Chinese large-scale DocumentVQA, spanning academic
papers, advertisements, web pages, and real-world photographed documents.                                  
\item We propose CDocRAG, a Chinese DocumentVQA system that improves retrieval accuracy by filtering
retrieved document images with a VLM-based relevance verifier before generation.
\item We conduct retrieval and gold-image generation experiments, demonstrating that CHaystack poses       
substantial challenges for current multimodal retrieval and generation models.                           
\end{itemize}

\section{Related Works}

\paragraph{\textbf{VQA benchmarks.}}

English VQA benchmarks have been extensively studied and now span a broad range of visual reasoning        
settings. Representative datasets include general VQA benchmarks such as VQA v2~\cite{vqa_v2}, together    
with benchmarks that target more specialized scenarios, including document understanding~\cite{docvqa},    
chart reasoning~\cite{chartqa}, mathematical reasoning~\cite{mathvista}, and retrieval-based               
VQA~\cite{retvqa}. Collectively, these datasets provide a relatively rich evaluation landscape for English
vision-language research.

In contrast, Chinese benchmarks in the DocumentVQA domain remain scarce. To the best of our knowledge, only
two representative Chinese benchmarks are currently available, namely
DuReader~\cite{qi_textrmdureader_textrmvis_2022} and MosaicDoc~\cite{mosaicdoc}. 
Both benchmarks, however, suffer from clear limitations. DuReader constitutes the first Chinese open-domain
DocumentVQA dataset and offers valuable large-scale supervision, yet its document sources remain largely
centered on web pages. MosaicDoc broadens the scope toward Chinese document understanding, but it primarily
focuses on newspapers and magazines and does not assess retrieval capability over a large candidate pool.

\paragraph{\textbf{Retrieval-augmented generation (RAG).}}
RAG was first developed in text-centric settings, where retrieval modules augment generation models with external knowledge~\cite{rag_original}. As multimodal foundation models continue to advance, RAG has gradually 
extended from pure-text tasks to settings that demand reasoning over both visual and textual evidence~\cite{hu_reveal_2023,jeong_videorag_2025}.

Nevertheless, existing multimodal RAG systems and benchmarks remain centered on English documents, while
Chinese large-scale document retrieval and question answering remain underexplored. Our work addresses this
gap by introducing CHaystack as a Chinese benchmark and CDocRAG as a corresponding system for    
large-scale document retrieval and answer generation. 

\begin{table}[t]
\centering
\caption{Comparison between CHaystack and existing Chinese DocVQA benchmarks.}
\label{tab:benchmark_comparison}
\small
\setlength{\tabcolsep}{3.5pt}
\begin{tabular}{lccccc}
\hline
\textbf{Benchmark} & \textbf{Retrieval} & \textbf{Language} & \textbf{Sources} & \textbf{Query} & \textbf{Docs} \\
\hline
DuReader    & \usym{2714} & CN & Web                              & 15K   & 128K  \\
MosaicDoc   & \usym{2718}   & CN, EN & Newspapers, Magazines                  & 622K  & 72K   \\
CHaystack   & \usym{2714} & CN & Papers, Ads, Web, Photo    & 4.5K  & 17.5K \\
\hline
\end{tabular}
\end{table}

\section{CHaystack Benchmark}
To support effective retrieval and reasoning across large-scale Chinese document collections, we present CHaystack, a new benchmark designed to evaluate whether DocumentVQA systems can locate the correct evidence document and generate the corresponding answer in diverse scenarios. Unlike existing Chinese benchmarks that mainly focus on a single document type, CHaystack organizes documents into four complementary categories: papers, web pages, advertisements, and photographed documents. This design enables a more systematic evaluation of Chinese DocumentVQA systems. CHaystack is constructed from multiple Chinese document and OCR-oriented resources, including CDLA for academic papers~\cite{buptlihang_buptlihangcdla_2026}, DuReader for web pages~\cite{qi_textrmdureader_textrmvis_2022}, XFUND and CC-OCR for photographed documents~\cite{xu_layoutxlm_2021, yang_cc-ocr_2024}, and MTWI for advertisements~\cite{he_icpr2018_2018}.
\begin{figure*}[htbp]
        \centering
        \includegraphics[width=\textwidth]{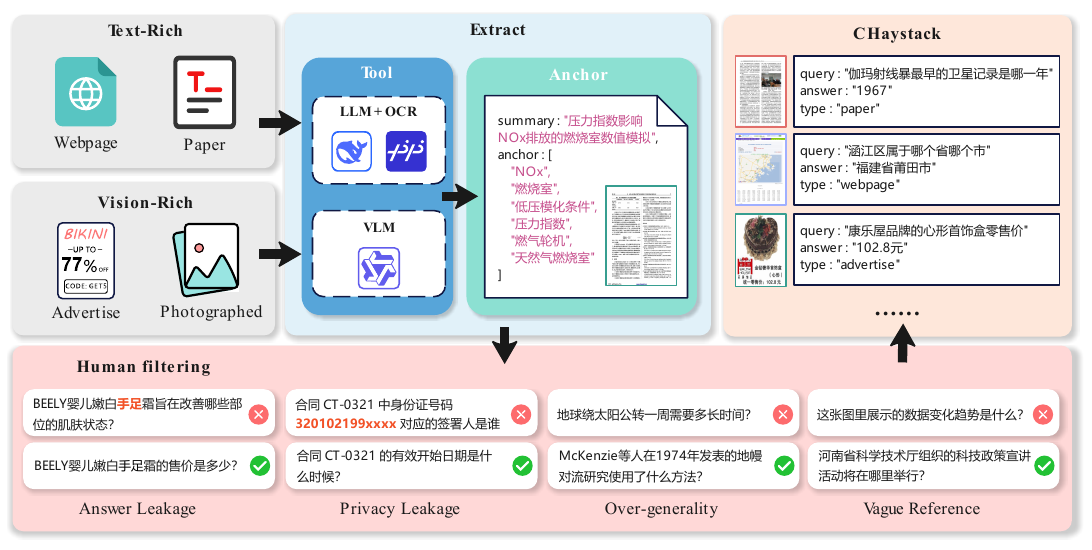}
        \caption{\textbf{Benchmark construction pipeline.} }
        \label{fig:benchmark}
\end{figure*}
\paragraph{\textbf{Benchmark construction pipeline.}}
As illustrated in Figure~\ref{fig:benchmark}, CHaystack is built through a construction pipeline based on anchors, which converts heterogeneous document images into query and answer pairs tied to specific documents. Since different document categories contain different types of useful evidence, we first divide the source documents into text rich and vision rich groups. For documents with rich text, including web pages and academic papers, we use DeepSeek-V3~\cite{deepseek_v3} together with PaddleOCR~\cite{paddleocr3} to extract textual anchors from OCR results. These anchors correspond to distinctive entities, phrases, numerical values, or local facts that can uniquely identify the source document. For documents with rich visual content, including advertisements and photographed documents, we use Qwen3-VL~\cite{qwen3_vl} to extract visual anchors from the image content, such as salient objects, layout cues, visual attributes, and associations between text and images that are difficult to capture using OCR alone.

After obtaining these anchors, we use both the anchor information and the original document image to generate query and answer pairs. Each query is designed to require evidence from a specific document, while the answer is grounded in the extracted anchor rather than generic world knowledge. This design reduces the risk of ambiguous questions that could be answered by multiple documents in the same collection. Finally, we combine automatic filtering by LLMs with human review to remove evaluation samples with low quality, including questions with vague references, answers that are not visually grounded, duplicate or overly general queries, and cases where multiple documents may lead to the same answer. Through this pipeline, CHaystack aims to provide reliable evaluation data tied to specific documents for Chinese DocumentVQA at large scale.
\paragraph{\textbf{Text-anchor construction.}}
For text-rich documents, we construct anchors from OCR text and cleaned document content. This part mainly covers academic papers and web pages, which contain dense text as well as distracting fragments. For academic papers, we retain evidence such as technical terms, institutional names, years, numerical values, and figure or table references, which help identify a specific paper page. For web pages, we first reduce boilerplate content such as navigation bars, advertisements, and side links, and then preserve central information such as page topics, named entities, dates, prices, phone numbers, and addresses. The goal is to keep anchors that are both distinctive and answerable, while removing fragments that are too common or unrelated to the main content. In this way, CHaystack converts noisy OCR outputs into compact document-specific evidence for large-scale retrieval.
\paragraph{\textbf{Visual-anchor construction.}}
For visually rich documents, we extract anchors that cannot be reliably captured by plain OCR alone. This part mainly covers advertisements and photographed documents, where evidence is often expressed through the combination of visible text, objects, layout, and image-specific attributes. For advertisements, we use anchors such as brand names, product information, salient slogans, numerical labels, promotional cues, and distinctive visual objects. For photographed documents, we use anchors such as document type, institution or company names, product or service fields, dates, amounts, and partial address information. These anchors are selected not as generic captions, but as discriminative cues for distinguishing the target document from visually or semantically similar candidates. This design encourages systems to use visual evidence for retrieval and reasoning, rather than relying only on common textual patterns.
\paragraph{\textbf{Question-answer generation and filtering.}}
Based on the extracted visual and textual anchors, we generate query-answer pairs for large-scale document retrieval and question answering. Each query must be self-contained and document-specific: it should include enough identifying information to retrieve the target document, but should not directly reveal the answer. Each answer must be short, unambiguous, and verifiable from the corresponding document image or OCR evidence. To reduce ambiguity, we filter out broad questions that may apply to many documents, questions with vague references such as ``this image'' or ``the above page'', and questions answerable from generic world knowledge. We also remove samples with answer leakage, weak grounding, unverifiable answers, duplicates, or privacy-sensitive information. If a question has clear evidence but is written in an unnatural form, it is rewritten into a fluent Chinese search query while preserving the original anchors and answer.
\begin{figure*}[t]
        \centering
        \includegraphics[width=0.85\textwidth]{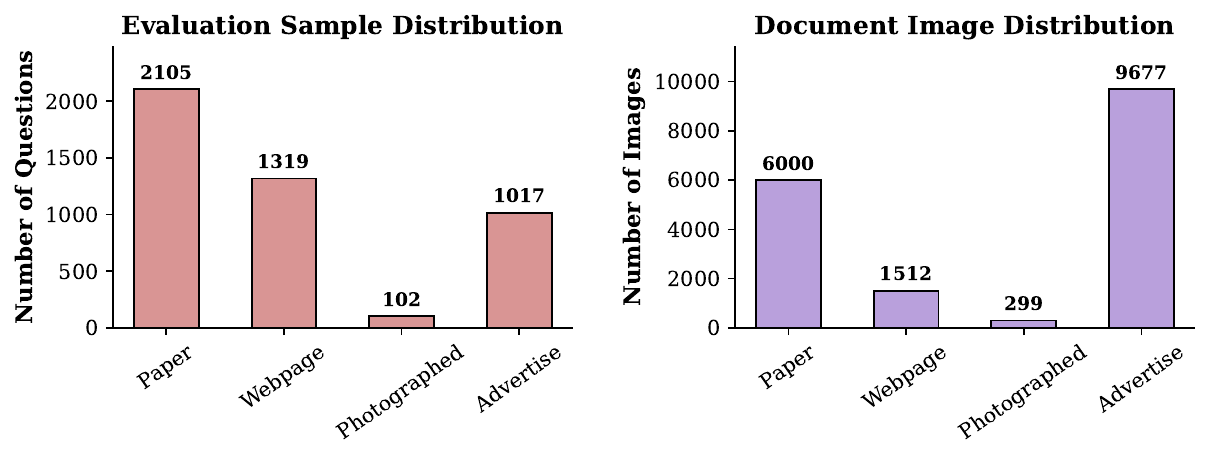}
        \caption{\textbf{Dataset profile of CHaystack.} We report the distribution of evaluation samples and document images across four document categories.}
        \label{fig:dataset_profile}
\end{figure*}
\paragraph{\textbf{Human filtering.}}
Beyond the automatic filtering steps described above, we further conduct a manual review to remove samples that fall into four problematic categories. \textbf{Answer leakage} refers to queries where the question formulation itself reveals or strongly implies the answer, making the question trivial to answer without consulting any document. \textbf{Privacy leakage} covers queries that expose personally identifiable information such as identity card numbers, phone numbers, or full names, which raises ethical concerns and should not appear in evaluation benchmarks. \textbf{Over-generality} describes questions that are answerable from common world knowledge and do not require access to a specific evidence document, thereby failing to test document-grounded reasoning. \textbf{Vague reference} concerns queries that rely on deictic expressions or underspecified context, such as ``this image'' or ``the above figure,'' making it impossible to identify the intended evidence document from the question alone. Through this multi-faceted human review, we ensure that each retained evaluation sample is document-specific, privacy-safe, and clearly answerable from its target evidence document.
\paragraph{\textbf{Final dataset profile.}}
As shown in Figure~\ref{fig:dataset_profile}, CHaystack contains 4,543 evaluation samples associated with 4,543 evidence documents. Specifically, the evaluation set includes 2,105 samples from academic papers, 1,319 from web pages, 1,017 from advertisements, and 102 from photographed documents. To assess retrieval performance at scale, these samples are evaluated over a larger candidate pool containing 17,488 document images, including 6,000 paper images, 1,512 web page images, 9,677 advertisement images, and 299 photographed document images. 

\section{Method}
We introduce \textbf{CDocRAG}, a filter-enhanced retrieval-augmented system for Chinese DocumentVQA. Instead of relying only on the ranked list produced by a multimodal retriever, CDocRAG first retrieves a relatively broad candidate set and then uses a vision-language model (VLM) as a relevance filter to judge whether each candidate image contains evidence for the question. Only the candidates that pass this filtering stage are passed to the generator for answer prediction. This design keeps the efficiency of embedding-based retrieval while adding an explicit evidence verification step before generation, making CDocRAG suitable for large-scale Chinese document retrieval and question answering.
\paragraph{\textbf{Initial retriever.}}
Given the original question, the retriever encodes it into a shared multimodal embedding space and searches over pre-encoded document image embeddings. We retrieve more candidates than the final generation budget, so that the following filtering stage has enough evidence candidates to examine. This stage provides high-throughput coarse localization and preserves the original retrieval scores and ranks for later analysis.
\paragraph{\textbf{VLM relevance filter.}}
The relevance filter takes the original question and each retrieved document image as input, and asks a VLM to make a binary judgment on whether the image contains sufficient evidence to answer the question. The filter is prompted to inspect visible text, layout, tables, figures, and other visual cues, and to output only \texttt{YES} or \texttt{NO}. Candidates judged as relevant are retained in their original order, while irrelevant candidates are removed. If all candidates are rejected, CDocRAG falls back to the original retrieved list to avoid producing an empty evidence set. The filter decisions are cached by question and image path, which allows the same filtering results to be reused across different generator runs.
\paragraph{\textbf{Generator.}}
The generator takes the original question together with the top-$k$ filtered document images as input and produces a concise answer. In our implementation, a Qwen-VL style multimodal generator is prompted to answer only according to the visible document content. If the filtered images do not contain sufficient evidence, the model is allowed to return an uncertain answer. The final output is normalized before evaluation, and the system reports both retrieval accuracy and answer quality.

\section{Experiments}

\subsection{Experimental Setup}

\paragraph{\textbf{Metrics.}}
For retrieval, we report Recall@$K$ with $K \in \{1,3,5,10\}$, which measures whether the gold evidence image appears in the top-$K$ retrieved candidates. For answer generation, we report exact match (EM) and character-level F1 after answer normalization, including punctuation removal, whitespace normalization, and lowercasing.

\paragraph{\textbf{Models.}}
For retrieval comparison, we evaluate representative retrieval models with fewer than 3B parameters, including Qwen3-VL-Embedding~\cite{qwen3_vl}, Chinese-CLIP~\cite{chinese_clip}, AltCLIP~\cite{altclip}, SigLIP~\cite{siglip}, and CLIP~\cite{clip}. For answer generation, we evaluate Qwen2.5-VL-3B-Instruct~\cite{qwen25_vl}, LLaVA-OneVision-Qwen2-0.5B~\cite{llava_onevision}, and InternVL2.5-4B~\cite{internvl25} under a gold-image setting, where the gold evidence image is directly provided to the model.

\paragraph{\textbf{Implementation details.}}
All experiments are conducted on the full CHaystack evaluation set, which contains 4,543 question-answer pairs and a retrieval candidate pool of 17,488 document images. For retrieval, we build category-specific image indexes and use the original question as the retrieval query. In the filter-based CDocRAG setting, we first retrieve 15 candidates, use Qwen2.5-VL-3B-Instruct as the relevance filter, and keep up to 10 filtered candidates for answer generation. For gold-image generation, the maximum generation length is set to 128 tokens.

\begin{table*}[ht]
\centering
\caption{Retrieval recall (R@$K$) on CHaystack by multimodal embedding model and document category.}
\label{tab:retrieval_recall}
\small
\setlength{\tabcolsep}{4pt}
\begin{tabular}{llcccc}
\hline
\textbf{Model} & \textbf{Category} & \textbf{R@1} & \textbf{R@3} & \textbf{R@5} & \textbf{R@10} \\
\hline
Qwen3-VL & All & \textbf{71.91} & \textbf{86.57} & \textbf{89.90} & \textbf{92.67} \\
 & Paper & 64.13 & 85.51 & 90.40 & 93.54 \\
 & Webpage & 85.75 & 92.57 & 93.63 & 94.69 \\
 & Advertisement & 68.93 & 80.73 & 83.87 & 88.20 \\
 & Photographed & 83.33 & 89.22 & 91.18 & 93.14 \\
\hline
Chinese-CLIP & All & 14.40 & 20.89 & 23.71 & 27.82 \\
 & Paper & 2.95 & 5.89 & 7.36 & 10.17 \\
 & Webpage & 6.29 & 10.01 & 11.98 & 15.77 \\
 & Advertisement & 47.49 & 64.31 & 70.80 & 77.19 \\
 & Photographed & 25.49 & 38.24 & 43.14 & 55.88 \\
\hline
AltCLIP & All & 11.05 & 16.93 & 19.68 & 23.38 \\
 & Paper & 2.47 & 4.94 & 5.99 & 7.60 \\
 & Webpage & 4.47 & 9.17 & 11.75 & 16.30 \\
 & Advertisement & 37.86 & 52.70 & 59.00 & 65.59 \\
 & Photographed & 5.88 & 7.84 & 12.75 & 19.61 \\
\hline
SigLIP & All & 1.63 & 2.47 & 3.02 & 3.96 \\
 & Paper & 0.19 & 0.24 & 0.29 & 0.76 \\
 & Webpage & 0.15 & 0.23 & 0.45 & 0.61 \\
 & Advertisement & 6.59 & 9.93 & 11.80 & 14.45 \\
 & Photographed & 0.98 & 2.94 & 4.90 & 8.82 \\
\hline
CLIP & All & 0.64 & 0.88 & 1.12 & 1.56 \\
 & Paper & 0.05 & 0.05 & 0.14 & 0.29 \\
 & Webpage & 0.38 & 0.61 & 0.76 & 1.36 \\
 & Advertisement & 2.26 & 2.85 & 3.44 & 4.13 \\
 & Photographed & 0.00 & 1.96 & 2.94 & 4.90 \\
\hline
\end{tabular}
\end{table*}

\begin{table*}[ht]
\centering
\caption{Gold-image answer generation on CHaystack by multimodal generator and document category. All models use the gold evidence image.}
\label{tab:gold_generation}
\small
\setlength{\tabcolsep}{4pt}
\begin{tabular}{llcc}
\hline
\textbf{Model} & \textbf{Category} & \textbf{EM} & \textbf{F1} \\
\hline
Qwen2.5-VL & All & \textbf{35.44} & \textbf{61.62} \\
 & Paper & 24.99 & 50.90 \\
 & Webpage & 46.17 & 72.64 \\
 & Advertisement & 39.92 & 66.95 \\
 & Photographed & 67.65 & 86.95 \\
\hline
LLaVA & All & 0.92 & 19.68 \\
 & Paper & 0.19 & 15.12 \\
 & Webpage & 0.68 & 23.16 \\
 & Advertisement & 2.46 & 22.79 \\
 & Photographed & 3.92 & 37.82 \\
\hline
InternVL2.5 & All & 20.89 & 42.64 \\
 & Paper & 20.76 & 42.05 \\
 & Webpage & 7.58 & 26.35 \\
 & Advertisement & 33.82 & 60.75 \\
 & Photographed & 66.67 & 84.98 \\
\hline
\end{tabular}
\end{table*}

\subsection{Main Experimental Results}

\paragraph{\textbf{Retrieval results.}}
Table~\ref{tab:retrieval_recall} shows that Qwen3-VL achieves the strongest retrieval performance among the compared models. On the full benchmark, it obtains 71.91 Recall@1, 86.57 Recall@3, 89.90 Recall@5, and 92.67 Recall@10. The advantage is especially clear on text-rich categories such as papers and webpages, where Qwen3-VL reaches 64.13 and 85.75 Recall@1, respectively.

Among the other compared models, Chinese-CLIP performs best overall, with 14.40 Recall@1 and 23.71 Recall@5, followed by AltCLIP with 11.05 Recall@1 and 19.68 Recall@5. Both models perform noticeably better on advertisements than on papers or webpages, indicating that contrastive image-text models can capture coarse visual semantics but struggle with dense document text and long-tail Chinese entities. In contrast, SigLIP and CLIP perform poorly across all categories, with full-set Recall@1 below 2\%. These results show that CHaystack is challenging for generic image-text encoders, especially in text-rich Chinese document scenarios.

\paragraph{\textbf{Gold-image generation results.}}
Table~\ref{tab:gold_generation} reports answer generation performance when the gold evidence image is provided. Qwen2.5-VL obtains the best overall result, achieving 35.44 EM and 61.62 F1. InternVL2.5 ranks second with 20.89 EM and 42.64 F1, while LLaVA-OneVision performs much worse, with only 0.92 EM and 19.68 F1. These results indicate that even when retrieval is solved, Chinese document question answering remains challenging: models must read fine-grained Chinese text, associate it with layout and visual context, and generate short answers in the expected form.

The category-level results further reveal different sources of difficulty. Photographed documents obtain the highest generation scores for Qwen2.5-VL and InternVL2.5, partly because many questions in this category are tied to salient document fields and short answers. Webpages and advertisements are more challenging but still benefit from strong multimodal reading ability. Academic papers are the most difficult category for Qwen2.5-VL, with 24.99 EM and 50.90 F1, reflecting the dense text, technical terminology, and structural complexity of paper pages. Overall, the retrieval and generation results together show that CHaystack evaluates both large-scale evidence localization and fine-grained Chinese document understanding, and that current open-source multimodal systems still leave substantial room for improvement.

\subsection{Ablation Study}

\begin{figure*}[t]
        \centering
        \includegraphics[width=0.65\textwidth]{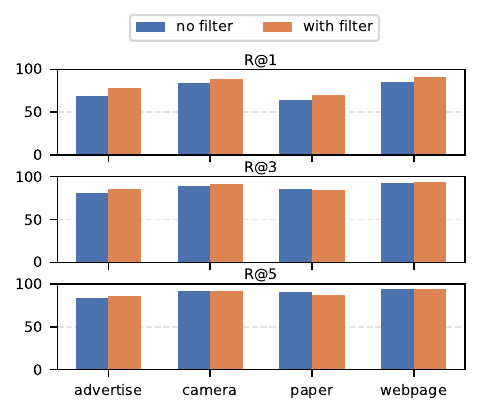}
        \caption{\textbf{Effect of the VLM relevance filter on retrieval recall.}}
        \label{fig:ablation_filter}
\end{figure*}

Figure~\ref{fig:ablation_filter} shows the effect of the relevance filter on retrieval recall, using Qwen3-VL-Embedding-2B as the retriever. On the full evaluation set, the filter improves Recall@1 from 71.82 to 78.32, with the largest gains on advertisements (68.83 to 78.47) and academic papers (64.04 to 70.07). At Recall@5, the filter still benefits advertisements and webpages but leads to a slight decrease on papers (90.36 to 86.46), due to false rejections of dense technical pages. Overall, the filter improves early precision at the cost of occasional false rejections on text-rich documents, which is an acceptable trade-off for downstream generation where only a small number of candidates are passed to the generator.

\section{Conclusion}

In this work, we introduced CHaystack, a Chinese DocumentVQA benchmark for evaluating retrieval and question answering across large-scale document images. CHaystack covers diverse document categories, including academic papers, webpages, advertisements, and photographed documents, providing a more realistic assessment of Chinese multimodal document understanding. To support this task, we proposed CDocRAG, a filter-enhanced retrieval-augmented framework that combines multimodal retrieval, VLM-based relevance filtering, and answer generation. Experimental results show that current models still face clear challenges in both dense Chinese document retrieval and fine-grained answer generation, indicating substantial room for future research on Chinese large-scale DocumentVQA.
%
%
%
\bibliographystyle{splncs04}
\bibliography{references}

@article{rag_original, author={Lewis, Patrick and Perez, Ethan and Piktus, Aleksandra and others}, title={Retrieval-augmented generation for knowledge-intensive NLP tasks}, journal={Advances in Neural Information Processing Systems}, volume={33}, pages={9459--9474}, year={2020}}

@article{zhang_comprehensive_2024, author={Zhang, Rui and Liu, Chen and Su, Yixin and others}, title={A Comprehensive Survey on Multimodal RAG}, journal={TechRxiv}, year={2024}}

@article{docvqa, author={Mathew, Minesh and Karatzas, Dimosthenis and Jawahar, CV}, title={DocVQA: A Dataset for VQA on Document Images}, journal={Proceedings of WACV}, pages={2200--2209}, year={2021}}

@misc{chen_document_2024, author={Chen, Jun and Xu, Dannong and Fei, Junjie and others}, title={Document Haystacks: Vision-Language Reasoning Over Piles of 1000+ Documents}, year={2024}}

@inproceedings{qi_textrmdureader_textrmvis_2022, author={Qi, Le and Lv, Shangwen and Li, Hongyu and others}, title={DuReader\_vis: A Chinese Dataset for Open-domain Document Visual Question Answering}, booktitle={Findings of ACL}, pages={1338--1351}, year={2022}}

@inproceedings{mosaicdoc, author={Chen, Ketong and Chen, Yuhao and Xue, Yang}, title={MosaicDoc: A Large-Scale Bilingual Benchmark for Visually Rich Document Understanding}, booktitle={Proceedings of AAAI}, pages={2913--2921}, year={2026}, doi={10.1609/aaai.v40i4.37282}}

@inproceedings{vqa_v2, author={Goyal, Yash and Khot, Tejas and Summers-Stay, Douglas and others}, title={Making the V in VQA Matter}, booktitle={Proceedings of CVPR}, pages={6904--6913}, year={2017}}

@inproceedings{chartqa, author={Masry, Ahmed and Long, Do Xuan and Tan, Jia Qing and others}, title={ChartQA: A Benchmark for Question Answering about Charts}, booktitle={Findings of ACL}, pages={2263--2279}, year={2022}}

@inproceedings{mathvista, author={Lu, Pan and Bansal, Hritik and Xia, Tony and others}, title={MathVista: Evaluating Mathematical Reasoning of Foundation Models in Visual Contexts}, booktitle={Proceedings of the International Conference on Learning Representations}, year={2024}}

@inproceedings{retvqa, author={Penamakuri, Abhirama Subramanyam and Gupta, Manish and Gupta, Mithun Das and Mishra, Anand}, title={Answer Mining from a Pool of Images: Towards Retrieval-Based Visual Question Answering}, booktitle={Proceedings of IJCAI}, pages={1312--1321}, year={2023}, doi={10.24963/ijcai.2023/146}}

@misc{hu_reveal_2023, author={Hu, Ziniu and Iscen, Ahmet and Sun, Chen and others}, title={REVEAL: Retrieval-Augmented Visual-Language Pre-Training with Multimodal Knowledge Memory}, year={2023}}

@misc{jeong_videorag_2025, author={Jeong, Soyeong and Kim, Kangsan and Baek, Jinheon and Hwang, Sung Ju}, title={VideoRAG: Retrieval-Augmented Generation over Video Corpus}, year={2025}}

@misc{buptlihang_buptlihangcdla_2026, author={buptlihang}, title={CDLA}, howpublished={\url{https://github.com/buptlihang/CDLA}}, year={2026}}

@misc{xu_layoutxlm_2021, author={Xu, Yiheng and Lv, Tengchao and Cui, Lei and others}, title={LayoutXLM: Multimodal Pre-training for Multilingual Visually-rich Document Understanding}, year={2021}}

@inproceedings{he_icpr2018_2018, author={He, Mengchao and Liu, Yuliang and Yang, Zhibo and others}, title={ICPR2018 Contest on Robust Reading for Multi-Type Web Images}, booktitle={Proceedings of ICPR}, pages={7--12}, year={2018}}

@misc{yang_cc-ocr_2024, author={Yang, Zhibo and Tang, Jun and Li, Zhaohai and others}, title={CC-OCR: A Comprehensive and Challenging OCR Benchmark}, year={2024}}

@misc{deepseek_v3, author={{DeepSeek-AI}}, title={DeepSeek-V3 Technical Report}, year={2024}}

@misc{paddleocr3, author={Cui, Cheng and others}, title={PaddleOCR 3.0 Technical Report}, year={2025}}

@misc{qwen3_vl, author={Bai, Shuai and others}, title={Qwen3-VL Technical Report}, year={2025}}

@inproceedings{clip, author={Radford, Alec and Kim, Jong Wook and Hallacy, Chris and others}, title={Learning Transferable Visual Models from Natural Language Supervision}, booktitle={Proceedings of ICML}, pages={8748--8763}, year={2021}}

@inproceedings{siglip, author={Zhai, Xiaohua and Mustafa, Basil and Kolesnikov, Alexander and Beyer, Lucas}, title={Sigmoid Loss for Language Image Pre-Training}, booktitle={Proceedings of ICCV}, pages={11975--11986}, year={2023}}

@article{chinese_clip, author={Yang, An and Pan, Junshu and Lin, Junyang and others}, title={Chinese CLIP: Contrastive Vision-Language Pretraining in Chinese}, journal={arXiv preprint arXiv:2211.01335}, year={2022}}

@inproceedings{altclip, author={Chen, Zhongzhi and Liu, Guang and Zhang, Bo-Wen and others}, title={AltCLIP: Altering the Language Encoder in CLIP}, booktitle={Findings of ACL}, pages={8666--8682}, year={2023}}

@article{qwen25_vl, author={Bai, Shuai and Chen, Keqin and Liu, Xuejing and others}, title={Qwen2.5-VL Technical Report}, journal={arXiv preprint arXiv:2502.13923}, year={2025}}

@article{llava_onevision, author={Li, Bo and Zhang, Yuanhan and Guo, Dong and others}, title={LLaVA-OneVision: Easy Visual Task Transfer}, journal={Transactions on Machine Learning Research}, volume={2025}, year={2025}}

@article{internvl25, author={Chen, Zhe and Wang, Weiyun and Cao, Yue and others}, title={InternVL2.5: Expanding Performance Boundaries of Open-Source Multimodal Models}, journal={arXiv preprint arXiv:2412.05271}, year={2024}}
\end{document}